\documentclass[twocolumn,showpacs]{revtex4}

\usepackage{amsmath}

\usepackage{graphicx}

\begin{document}
\draft
\title{Glass-like low frequency ac response of ZrB$_{12}$
and Nb single crystals in the surface superconducting state}

\author{Menachem I. Tsindlekht, Grigory I. Leviev, Valery M. Genkin, and Israel Felner}
\affiliation{The Racah Institute of Physics, The Hebrew University
of Jerusalem, 91904 Jerusalem, Israel}
\author{Yurii B. Paderno and  Vladimir B. Filippov}
\affiliation{Institute for Problems of Materials Science, National
Academy of Sciences of Ukraine, 03680 Kiev, Ukraine}

\begin{abstract}
We report experimental studies of the low frequency electrodynamics
of ZrB$_{12}$ and Nb single crystals. AC susceptibility at
frequencies 3 - 1000 Hz have been measured under a dc magnetic
field, $H_0$, applied parallel to the sample surface. In the surface
superconducting state for several $H_0$ the real part of the ac
magnetic susceptibility exhibits a logarithmic frequency dependence
as for spin-glass systems. Kramers-Kronig analysis of  the
experimental data, shows  large losses at ultra low frequencies
($<3$ Hz). The wave function slope at the surface was found. The
linear response of the order parameter to the ac excitation was
extracted from the experimental data.

\end{abstract}

\date{\today}
\maketitle

\section{Introduction}

Nucleation of the superconducting phase in a thin surface sheath in
a decreasing magnetic field parallel to the sample surface was
predicted by Saint-James and de Gennes some years ago~\cite{PG}.
They showed that nucleation of the superconducting phase occurs in a
magnetic field $H_0<H_{c3} = 2.39~\kappa H_c$, where $H_c$ is the
thermodynamic critical field and $\kappa$ is the Ginzburg-Landau
(GL) parameter. Experimental confirmations of this prediction were
done in numbers of publications~\cite{STR,PAS,BURG,ROLL,SWR}. It was
found that in the in surface superconducting state (SSS) at low
frequencies, the superconducting compounds are lossy  materials.
However, the losses in the SSS, which exceed the losses of the
normal state, shows a peak at external fields $H_0$, when
$H_{c2}<H_0<H_{c3}$. This peak exists even at frequencies of a few
Hertz. In general, the response is nonlinear, and frequency
dependent. The critical state model was found adequate for the
description of the experimental data~\cite{ROLL}.

In the last few years the SSS has attracted renewed interest from
various directions~\cite{TS2,JUR,GM,SCOL,RYDH,LEV2,FINK1}. The
stochastic resonance phenomena in the  SSS for Nb single-crystal,
were observed in the nonlinear low-frequency response to ac fields,
~\cite{TS2}. In~\cite{JUR} it was assumed that at $H_0=H_{c3}$ the
sample surface consists of  many disconnected superconducting
clusters and subsequently the percolation transition takes place at
$H_{c3}^{c}=0.81 H_{c3}$. The paramagnetic Meissner effect is also
related to the SSS ~\cite{GM}. Voltage noise and surface current
fluctuations in Nb in the SSS has been investigated ~\cite{SCOL}.
Surface superconducting states were detected also in single crystals
of MgB$_2$~\cite{RYDH} and ZrB$_{12}$~\cite{LEV2}. A new theoretical
approach, based on a generalized form of the GL functional, was
developed in~\cite{FINK1}. Surface superconductivity in three
dimensions was studied theoretically in~\cite{XN,MOG}. It was found
~\cite{ROLL,LEV2} that, in general, the wave form of the surface
current in an ac magnetic field has a non-sinusoidal character. The
latter one can be described by a simple phenomenological relaxation
equation for transitions between metastable states~\cite{LEV2}. The
relaxation time in this equation depends on the deviation from
equilibrium and increases with decreasing of the excitation
frequency~\cite{LEV2}.

In spite of the extensive study, the origin of low frequency losses
in SSS under weak ac fields, is not clear yet. The critical state
model implies that if the amplitude of the ac field is smaller than
some critical value, the losses are absent. Indeed, the experimental
results for Pb-2$\%$In alloy confirm this prediction~\cite{ROLL}. On
the other hand, the observed response ~\cite{JUR} for an excitation
amplitude of 0.01 Oe (that, is considerably smaller than that used
in~\cite{ROLL}), shows losses in SSS of Nb at 10 Hz. Our measurement
on Nb single crystal at 733 Hz also has shown that the out-of-phase
part of the ac susceptibility, $\chi^{''}$, is independent of the
amplitude, and is finite at low amplitude values. We consider these
results as confirming that the losses in SSS are caused by
essentially linear phenomena and thus the critical state model
cannot be used for an adequate description of the ac response.

In this paper we present detailed experimental study of the linear
low-frequency response of ZrB$_{12}$ ($T_c=6.06$~K,
$\kappa\approx0.75$) and Nb ($T_c=9.2$ K, $\kappa\approx1.5$) single
crystals in the SSS, for frequencies $3\leq\omega/2\pi\leq 1000$~Hz.
We show that already at 3 Hz the ac susceptibility does not equal
$\partial M/\partial H_0$, where $M$ is the dc equilibrium
magnetization. For some dc magnetic fields $\chi^{'}\propto
\ln\omega$, which resembles spin-glass systems ~\cite{SG}. The
Kramers-Kronig analysis of our experimental data predicts that for
several dc magnetic fields huge loss peak should exist at very low
frequencies. It is believed that the observed response presents the
average value over many clusters, each of which is governed by a
second order differential equation, with individual relaxation
parameters. This is unlike a spin-glass system where the first order
differential equation can be used~\cite{LUN}. The observed ratio of
$H_{c3}/H_{c2}$ for our crystals differs from the predicted value -
1.69. It is believed that this is due to the nonzero slope of the
wave function at the surface,
$b=-\frac{1}{\Psi}\frac{\partial\Psi}{\partial x}$. The increased
ratio of $H_{c3}/H_{c2}$ shows that the sign of $b$ is unexpectedly
negative~\cite{ANG}.

The order parameter, $\Psi$, in the SSS has the form:
 \begin{equation}\label{Eq1}
   \Psi(x,y) = f(x)exp(iKy),
\end{equation}
where $f(x)$  equals zero inside the bulk of the sample ($x$-axis is
assumed normal to the sample surface, $H_0$ parallel to $z$-axis)
and $K$ is some as yet undetermined constant. In general case for
any dc magnetic field there is a band of possible values for $K$ for
which the solution of the GL equations can be found. These solutions
describe the metastable surface states with nonzero total surface
current and only the solution with zero current corresponds to the
equilibrium state. In this state the dc magnetic moment equals zero
for $H>H_{c2}$. On the other hand, in the presence of an ac field,
the nonzero response in SSS shows that the ac moment is not zero,
thus the system is not in equilibrium. In the linear approximation
\begin{equation}\label{Eq2}
   K(t) = K_0+\text{Re}[G(\omega,H_0)h(\omega)\text{exp}(-i\omega
   t)],
\end{equation}
where the external magnetic field is
$H(t)=H_0+\text{Re}[h(\omega)\text{exp}(-i\omega t)]$. The function
$G(\omega,H_0)$ characterizes the response of the order parameter to
an ac field and is determined in this paper.

\section{Experimental Details}

The measurements were carried out on ZrB$_{12}$ and Nb single
crystals, which were grown in the Institute for Problems of
Materials Science NAS, Ukraine, and in the Institute of Solid State
Physics RAS, Russia, respectively. The dimensions of the crystals
are $10.3\times 3.2\times 1.2$ mm$^3$ for ZrB$_{12}$ and $10\times
3\times 1$ mm$^3$ for Nb. The details of the sample preparation and
their magnetic characteristics were published
previously~\cite{TS1,TS2}. All the magnetization curves were
measured using a SQUID magnetometer. Ac susceptibility, in-phase,
$\chi^{'}$, and out-of-phase, $\chi^{''}$, components were measured
using the pick-up coils method~\cite{SH,ROLL}. Each sample was
inserted into one of a balanced pair coils. The unbalanced signal as
a function of the external parameters: temperature, dc magnetic
field, frequency and amplitude of excitation, was measured by a
lock-in amplifier. The experiment was carried out as follows. The
crystal was cooled down at zero magnetic field (ZFC). Then the
magnetic field was applied. The amplitude and the phase of the
unbalanced signal were measured in a given magnetic field, including
zero field, at all frequencies. The amplitude of excitation was
$0.01\div0.5$ Oe. We have supposed that in zero dc magnetic field
the ac crystal susceptibility is equal to the dc susceptibility in
the Meissner state with negligible losses. It permits us to find an
absolute value of the in-phase and out-of-phase components of the ac
magnetic susceptibility for all applied fields and frequencies. A
"home-made" measurement cell of the experimental setup was adapted
to a commercial SQUID magnetometer. The block diagram of the
experimental setup has been published elsewhere~\cite{LEV2}.

For the Fourier component of the magnetization
$m(t)=\text{Re}[m(\omega,H_0,h(\omega))\text{exp}(-i\omega t)]$ in
the linear approximation one can write
\begin{equation}\label{Eq3}
\begin{array}{c}
 m(\omega,H_0,h(\omega))=\chi_n(h(\omega)+4\pi J_s(\omega,H_0,h(\omega))/c)+\\
\bigskip
4\pi\chi_sJ_s(\omega,H_0,h(\omega))/c,
\end{array}
\end{equation}
where $\chi_n$ is the susceptibility of the normal core of the
sample, and $J_s(\omega,H_0,h(\omega))$ is the Fourier component of
the surface supercurrent,
$J_s(t)=\text{Re}(J_s(\omega,H_0,h(\omega)))\text{exp}(-i\omega t)$.
This equation takes into account the magnetic moment of the normal
core of the bulk and the magnetic moment of the surface
supercurrent~\cite{ROLL}. With Eq.~(\ref{Eq3}) we can find
$J_s(\omega,H_0,h(\omega))$ and the "surface susceptibility" defined
as $\chi_s(\omega,H_0)\equiv J_s(\omega,H_0,h(\omega))/h(\omega)$.
The measured  susceptibility , $\chi(\omega,H_0)\equiv
m(\omega,H_0,h(\omega))/h(\omega)$ and "surface susceptibility",
$\chi_s$, are connected to each other as follows:

\begin{equation}\label{Eq4}
  \chi_s(\omega,H_0)=[\chi(\omega,H_0)-\chi(\omega,H_n)]/[1+4\pi\chi(\omega,H_n)],
\end{equation}
where $H_n>H_{c3}$ is the magnetic field at which the sample is in
the normal state. These quantities, ( $J_s$ and $\chi_s$),
characterize the response of the surface current and eliminate the
contribution of the normal core in the bulk and the small unbalanced
signal of the empty coils

\section{Experimental results}

\begin{figure}
     \begin{center}
    \leavevmode
       \includegraphics[width=0.9\linewidth]{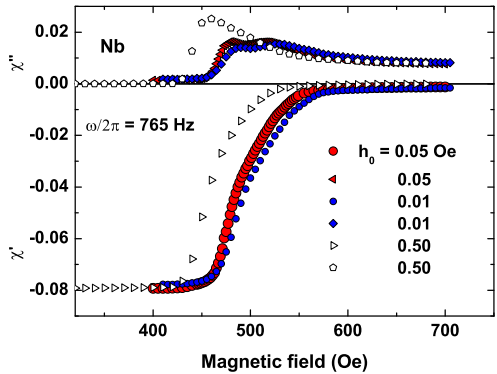}

        \bigskip

    \caption{(Color online) Magnetic field dependencies of
    $\chi^{'}$ and
    $\chi^{''}$
        for Nb at $T=8.5$~K at different amplitudes of excitation, $h_0$.}

     \label{f-1a}
     \end{center}
     \end{figure}

The field dependencies of the ac susceptibility $\chi^{'}$,
$\chi^{''}$ of Nb single crystal at different excitation amplitudes
($h_0$) are presented in Fig.~\ref{f-1a}. The data show clearly that
at the low amplitudes $\chi^{'}$, $\chi^{''}$ are almost independent
on the $h_0$. Similar results were obtained for the ZrB$_{12}$
crystal. We consider that at $h_0\leq 0.05$~Oe the observed response
in both Nb and ZrB$_{12}$ crystals has a linear origin.

\begin{figure}
     \begin{center}
    \leavevmode
       \includegraphics[width=0.9\linewidth]{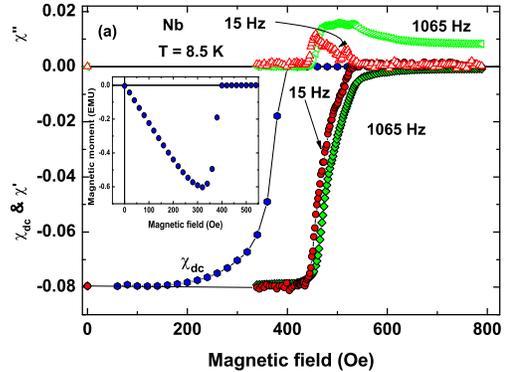}
       \includegraphics[width=0.9\linewidth]{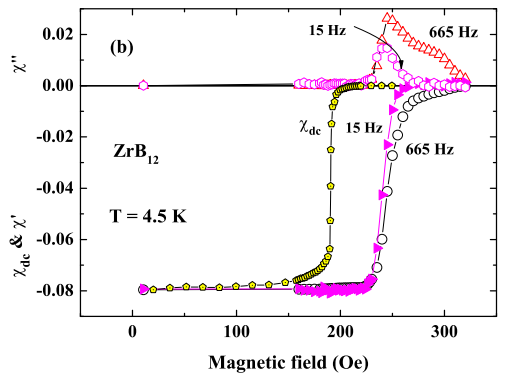}
        \bigskip

       \caption{(Color online) (a) Magnetic field dependencies of $\chi^{'}$, $\chi^{''}$
        and $\chi_{dc}=M/H_0$ for Nb at $T=8.5$~K.\\
        (b) Magnetic field dependencies of $\chi{'}$, $\chi{''}$ and
        $\chi_{dc}=M/H_0$ for ZrB$_{12}$ at $T=4.5$~K.}
     \label{f-1}
     \end{center}
     \end{figure}

Figs.~\ref{f-1}a and Fig.~\ref{f-1}b show the $\chi^{'}$,
$\chi^{''}$ and the ZFC dc susceptibility, $\chi_{dc}\equiv M/H_0$,
as a function of the dc field ($H_0$) of Nb and ZrB$_{12}$
respectively. The insets in Figs.~\ref{f-1}a and~\ref{f-1}b present
the ZFC isothermal magnetization curve.
Note, that for both crystals, at $H_0>H_{c2}$ the dc susceptibility
in the SSS is zero, whereas a large and diamagnetic ac
susceptibility signal is observed, see Fig.~\ref{f-1}. The
difference between the dc and the in-phase ac susceptibility signals
remain even at $\omega/2\pi=3$~Hz (the lowest frequency in our set
up). The out-of-phase component, $\chi^{''}$, has a broad maximum
for $H_0>H_{c2}$.

In the normal state, for $H_0 >H_{c3}$, the losses for Nb are
greater than for ZrB$_{12}$ (see Fig. 2a for 1065 Hz), because the
normal state conductivity of Nb is larger than that of ZrB$_{12}$.
The out-of-phase susceptibility $\chi^{''}$ is proportional to
$\omega$ as expected for the normal skin effect in the limit
$\delta>>d$, where $\delta$ is the skin depth and $d$ is the sample
thickness~\cite{LL}. For both crystals, when $H_0>H_{c3}$,
$\chi{''}$ is field independent. However in SSS, (see
Fig.~\ref{f-1}), the losses for Nb are lower than for ZrB$_{12}$.
Both, the in- and out-of-phase components of the ac susceptibility,
$\chi(\omega,H_0)$, do not show any clear peculiarities as the dc
magnetic field passes through $H_{c3}$. Therefore, it is difficult
to extract the  $H_{c3}$ value from the experimental data. More
sensitive measurement of $H_{c3}$ can be done by using the
$\chi_s(\omega,H_0)$ curves (see below).

\begin{figure}
     \begin{center}
    \leavevmode
       \includegraphics[width=0.9\linewidth]{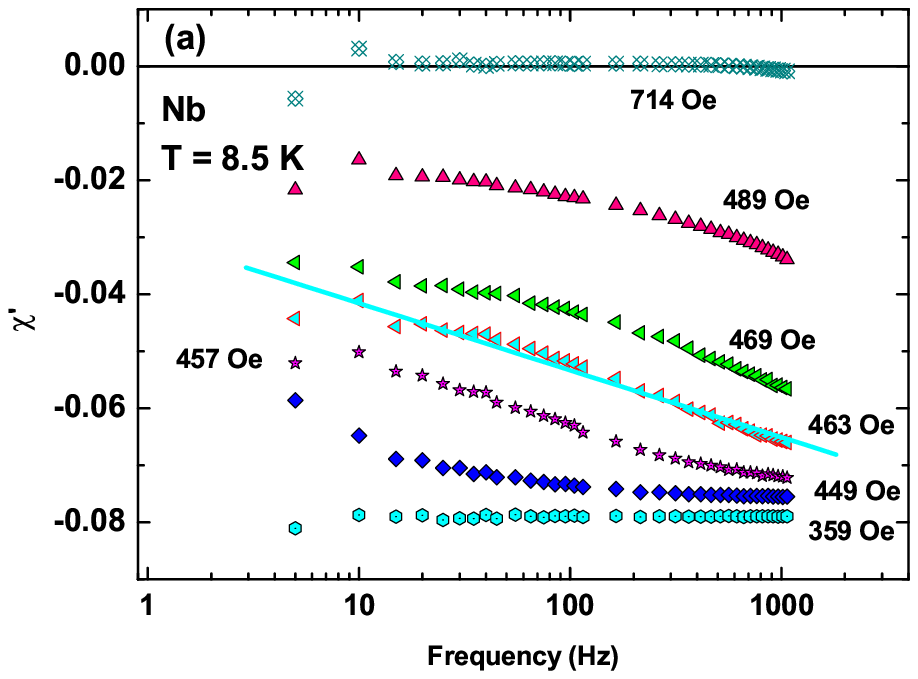}
       \includegraphics[width=0.9\linewidth]{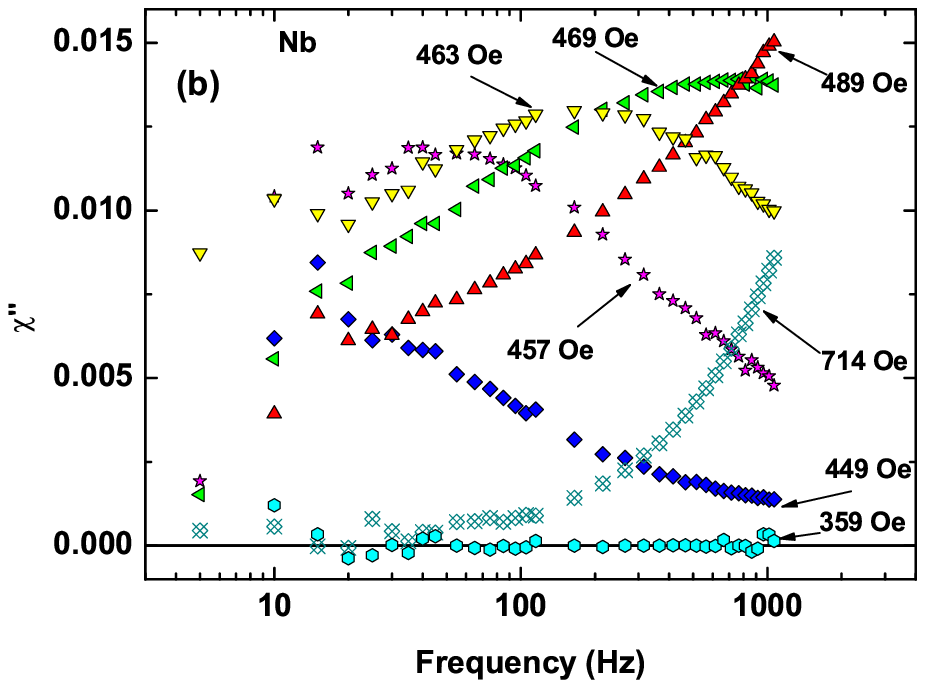}

    \caption{(Color online) Frequency dependence of the ac susceptibility for Nb at
       $T=8.5$~K at dc magnetic field $H_0>H_{c2}$. (a) - $\chi^{'}(\omega)$  and (b) -
        $\chi^{''}(\omega)$.}

     \label{f-2}
     \end{center}
     \end{figure}

\begin{figure}
     \begin{center}
    \leavevmode
       \includegraphics[width=0.9\linewidth]{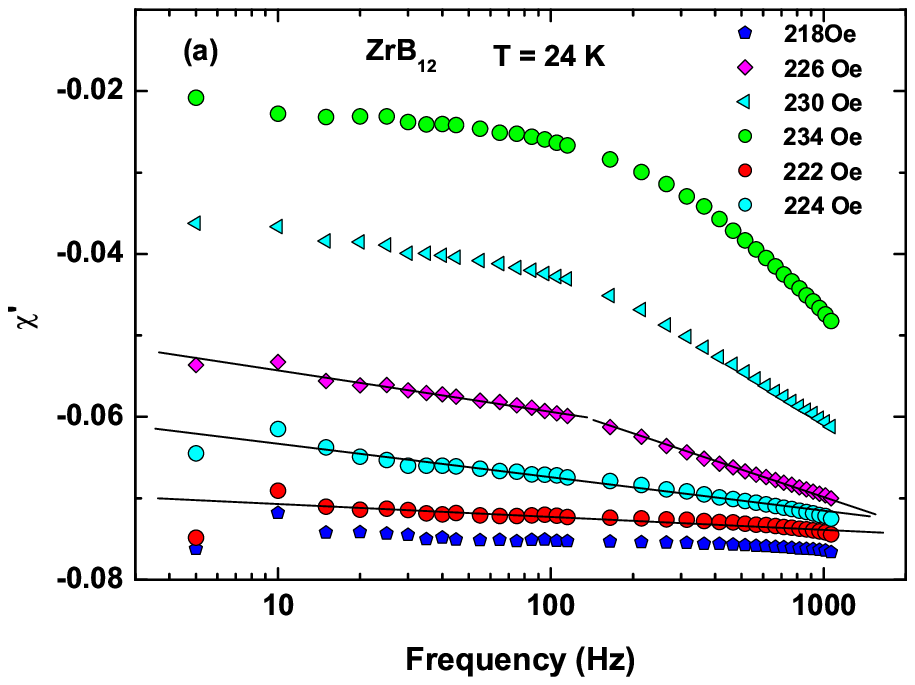}
       \includegraphics[width=0.9\linewidth]{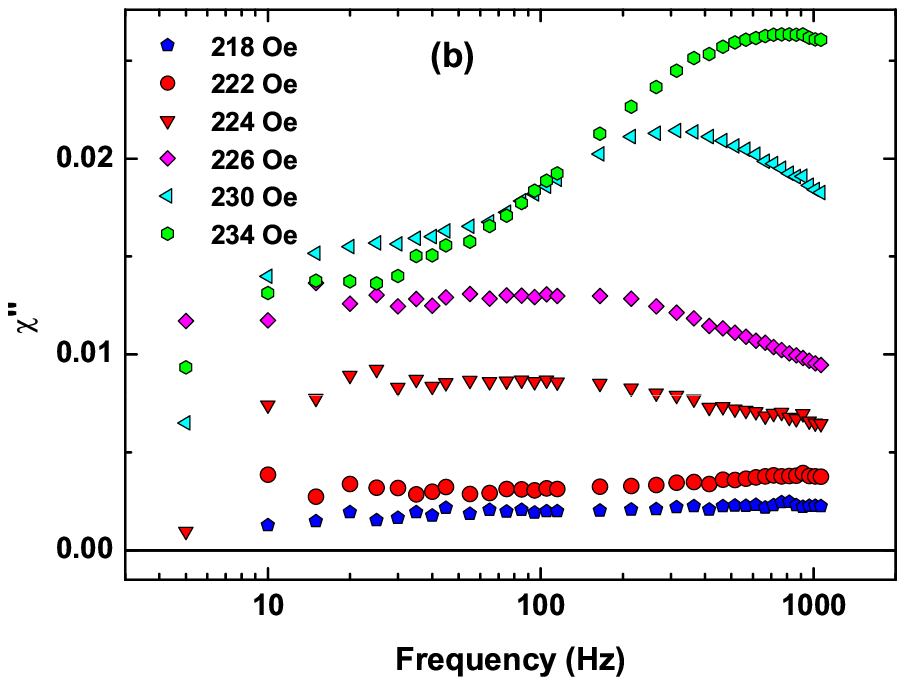}
        \bigskip

       \caption{(Color online) Frequency dependence of the ac susceptibility for ZrB$_{12}$ at
       $T=8.5$~K at dc magnetic field $H_0>H_{c2}$. (a) - $\chi^{'}(\omega)$  and (b) -
        $\chi^{''}(\omega)$.}

     \label{f-3}
     \end{center}
     \end{figure}

\begin{figure}
     \begin{center}
    \leavevmode
       \includegraphics[width=0.9\linewidth]{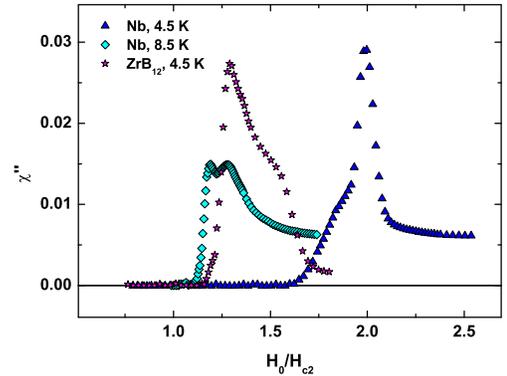}
    \caption{(Color online) The field dependence of $\chi^{''}$
         for Nb at $T=4.5$~K, 8.5~K and for ZrB$_{12}$ at $T=4.5$~K
         (where $\omega/2\pi=765$ Hz).}

     \label{f-5}
     \end{center}
     \end{figure}

Figs.~\ref{f-2} and~\ref{f-3} demonstrate the frequency dependence
of $\chi^{'}$ and $\chi^{''}$ for both Nb and ZrB$_{12}$ crystals at
dc magnetic field above $H_{c2}$. The character of these curves
depends on the field. Thus $\chi^{'}$ is a concave function of
$\omega$ at fields close to $H_{c2}$ and is a convex function close
to $H_{c3}$. For some fields, $\chi^{'}\propto\ln\omega$ for both
samples. Such a logarithmical dependence is typical of spin glasses
which are completely different physical systems ~\cite{SG}.
Fig.~\ref{f-5} presents $\chi^{''}$ as a function of the reduced
magnetic field, $H_0/H_{c2}$ for two frequencies and temperatures
for Nb and ZrB$_{12}$ crystals. The $H_{c3}/H_{c2}$ ratio is
temperature dependent as previously reported for Nb (e.g.
~\cite{HOP}).

\section{ Theoretical background }

\begin{figure}
     \begin{center}
    \leavevmode
       \includegraphics[width=0.9\linewidth]{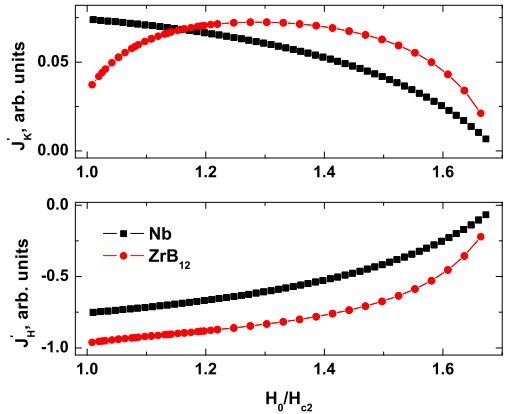}

    \caption{(Color online) The derivatives of surface current $J_H^{'}$
 $J_K^{'}$ versus reduced magnetic field for Nb and ZrB$_{12}$ (see text).}

     \label{f-6}
     \end{center}
     \end{figure}

For interpretation of the experimental data we used the
numerical approach to the normalized stationary GL equations
\begin{equation}\label{Eq5}
  \begin{array}{c}
    -(i{\nabla}/\kappa+\overrightarrow{A})^2\Psi^2+\Psi-\mid\Psi\mid^2\Psi=0\\
   -\text{curl}\text{curl}\overrightarrow{A}=\overrightarrow{A}\mid\Psi\mid^2+ i/2\kappa(\Psi^*\nabla\Psi-\Psi\nabla\Psi^*)\\
  \end{array}
\end{equation}

for an external magnetic field parallel to the sample surface. The
order parameter, $\Psi$, is normalized with respect to its value at
zero magnetic field, the distances with respect to the London
penetration length $\lambda$, and the vector potential
$\overrightarrow{A}$ with respect to $\sqrt 2 H_c\lambda$, where
$H_c$ is the thermodynamic critical field. Assuming that the order
parameter has the form of Eq.~(\ref{Eq1}) with the yet as
undetermined parameter $K$, we see that $K$ is the integral constant
of these equations, if the sample thickness considerably exceeds the
coherence length and the superconductor is homogeneous. The
relaxation time in the time-dependent version of GL theory is of the
order of $10^{-11}-10^{ -13}$~s and we can use the stationary
version of the GL theory, Eq.~(\ref{Eq5}), for the ac experiments.
Equations~(\ref{Eq5}) with proper boundary conditions and the
requirement that the order parameter differs from zero only near the
surface, have the solution for a whole band of $K$ values. But only
one value of $K$ corresponds to the total surface current equals
zero and this $K$ describes the equilibrium state with the minimal
free energy. These nonlinear equilibrium surface solutions have been
discussed in detail in~\cite{FINK}. The ac response for
superconductors in fields $H_{c2}<H_0<H_{c3}$ differs from the
normal one. This means that in an ac field the superconductor is in
a nonequilibrium state with finite surface current and
nonequilibrium $K$. Generally speaking, the total surface current
depends on both: the instant values of external magnetic field
$H(t)$ and on $K(t)$,thus $J_s(t)=J_s(H(t),K(t))$. If a small ac
magnetic field $h(t)$ is superimposed upon a dc field $H_0$, the
amplitude of the surface current in the linear approximation is

\begin{equation}\label{Eq6}
\begin{array}{c}
J_s(\omega,H_0,h(\omega))=\frac{\partial J_s(H_0,K_0)}{\partial
H_0}h(\omega)\\+\frac{\partial J_s(H_0,K_0)}{\partial
K_0}G(\omega,H_0)h(\omega).
\end{array}
\end{equation}
where $K_0$ is the equilibrium value of $K$ in a dc magnetic field
$H_0$, and $G(\omega,H_0)$ describes the linear response of $K$ to
an ac field Eq.~(\ref{Eq2}). The partial derivatives in
Eq.~(\ref{Eq6}) ($J^{'}_K\equiv\frac{\partial J_s(H_0,K_0)}{\partial
K_0}$ and $J^{'}_H\equiv\frac{\partial J_s(H_0,K_0)}{\partial H_0}$)
can be calculated numerically and in Fig.~\ref{f-6} we show the
results for Nb (GL parameter $\kappa=1.5$~\cite{HOP}) and ZrB$_{12}$
($\kappa=0.75$~\cite{TS1}).

Equation~(\ref{Eq6}) describes the linear response of the surface
current to an external ac magnetic field. If $K$ does not change
during an ac cycle, we see from Fig.~\ref{f-6} that only a smooth
decrease of the surface current without any losses should be
observed, as the dc magnetic field increases. The critical current
model assumes that for $h(\omega)<h_c$, where $h_c$ is some critical
ac field, the surface current follows the external magnetic field
without any delay, $\text{Im}G(\omega,H_0)=0$, and therefore surface
losses are absent.

\section{Discussion}

\begin{figure}
     \begin{center}
    \leavevmode
       \includegraphics[width=0.9\linewidth]{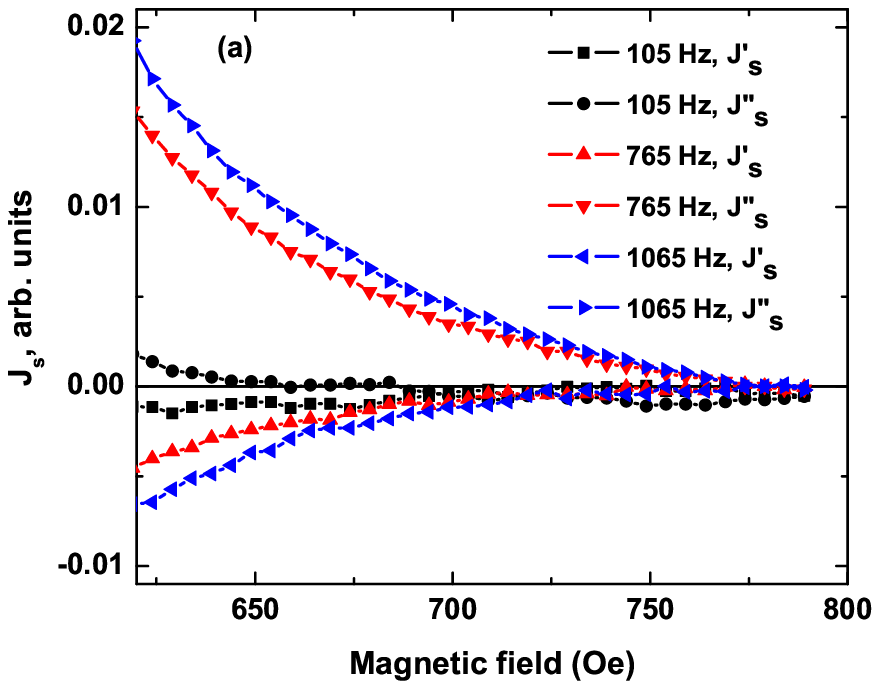}
        \includegraphics[width=0.9\linewidth]{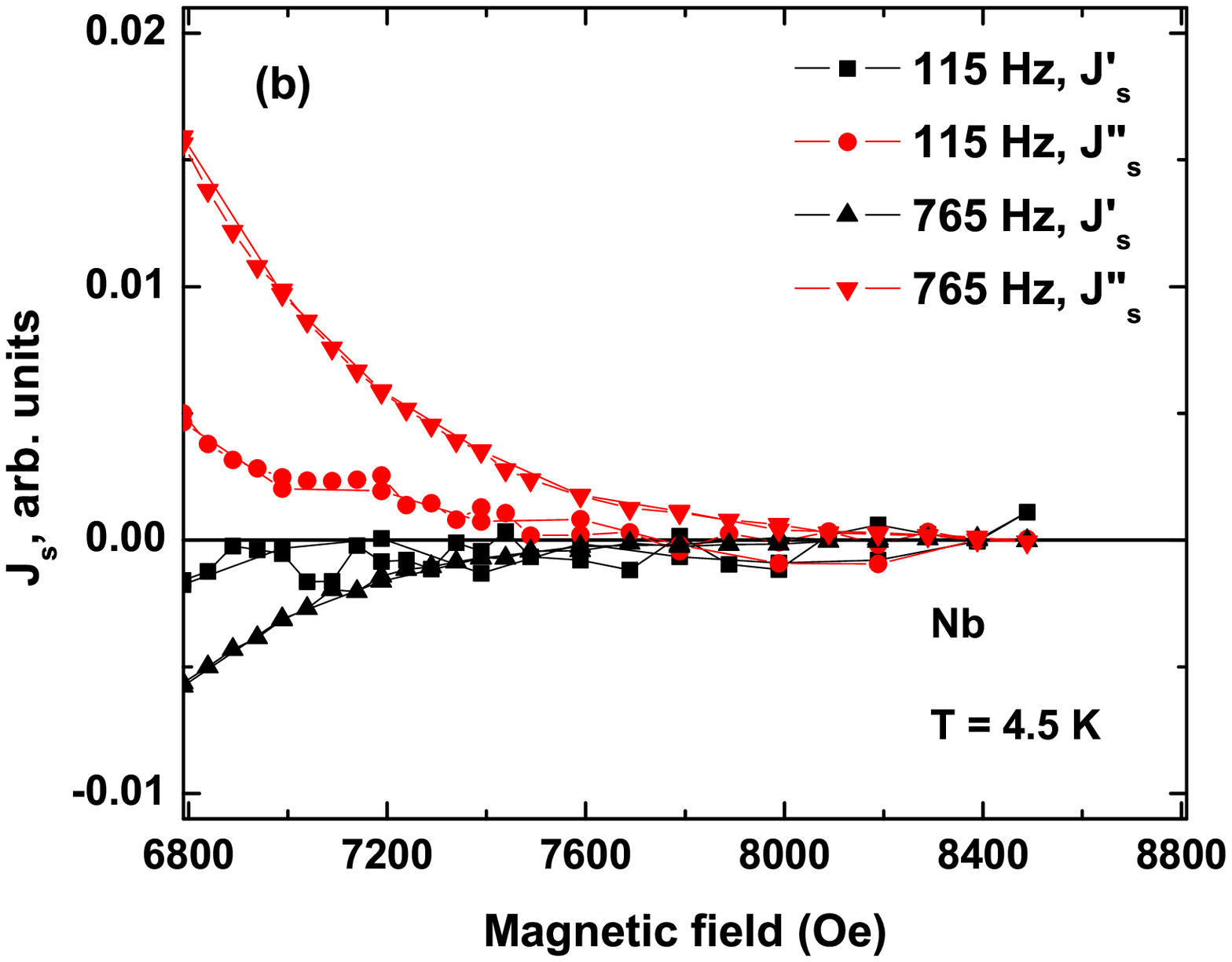}
    \caption{(Color online) The field dependence
    of the real, $J_s^{'}$, and imaginary, $J_s^{''}$, parts of surface current
      at different frequencies
         for Nb at T=8.5~K , $H_{c2}=400$ Oe (a); 4.5~K, $H_{c2}=3200$ Oe (b).}

     \label{f-7}
     \end{center}
     \end{figure}

\begin{figure}
     \begin{center}
    \leavevmode
       \includegraphics[width=0.9\linewidth]{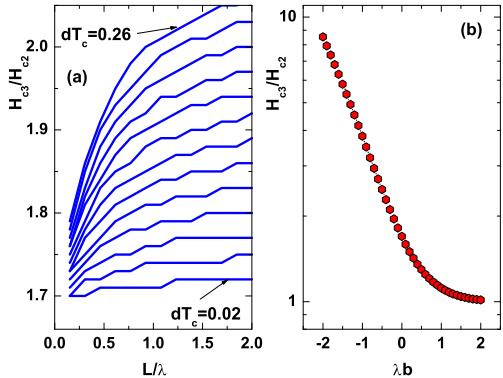}
    \caption{(Color online) The  $H_{c3}/H_{c2}$ ratio for two
    cases:\\
    (a) $T_c$ in the surface layer (of thickness L) is higher than in the bulk value
    see Eq.~(\ref{Eq7}));\\
    (b) the slope of the wave function,
     $b$, at the surface is different from zero ($\kappa=1.5$).}

     \label{f-8}
     \end{center}
     \end{figure}

It is clear that the experimental data show the existence of SSS in
our crystals. Usually $H_{c3}$ is defined by the onset of the
surface screening, i. e. by the appearance of the deviation of the
ac response from its normal value, as the dc field decreases from
some large value. Because in the SSS the ac response is frequency
dependent, using of low frequencies could give the underestimated
value of $H_{c3}$. As an example, in Fig.~\ref{f-7}a we show a set
of data obtained for Nb sample at 8.5 K, at frequencies 115, and 765
Hz.
The imaginary part of surface current is more sensitive for $H_{c3}$
determination (obviously, the experimental setup sensitivity is
higher at higher frequencies). The data at a frequency of about 100
Hz yield $H_{c3}\cong 680$~Oe, while at higher frequencies ($\approx
700$ Hz) we obtain $H_{c3}\cong 760$~Oe, in the later case
$H_{c3}/H_{c2}=1.9$. This value is considerably larger than the
value predicted 1.69 in Ref.~\cite{PG} With decreasing the
temperature, the discrepancy between the theoretical
prediction~\cite{PG} and the experimental values increases and the
ratio at $T=4.5$~K becomes $H_{c3}/H_{c2}=2.34$ (Fig.~\ref{f-7}b).
The decrease of this ratio for temperatures in the vicinity of
$T_c$, was found in several experiments~\cite{HOP,OST} and it was
associated with the decrease of $T_c$ near the surface~\cite{HU}. In
the framework of the GL theory the $H_{c3}/H_{c2}$ ratio can be
changed only if (i) either  $T_c$ in the surface layer differs from
the bulk value, or (ii) the slope of the wave function at the
surface $b=-\frac{1}{\Psi}\frac{\partial \Psi}{\partial x}$ differs
from zero~\cite{ANG,JOI}. In Fig.~\ref{f-8} we show the calculated
$H_{c3}/H_{c2}$ ($\kappa=1.5$) for these two cases.

 The boundary condition for the
 first case is $b=0$ but the $T_c$ values of a surface layer
with thickness $L$, increases as:
\begin{equation}\label{Eq7}
   (T_c(x)-T_{c})/(T_{c}-T)=dT_c\text{exp}(-x/L)
\end{equation}
(Fig.~\ref{f-8}a). In the second one,(when $dT_c=0$), we assume that
$b\neq 0$ (Fig.~\ref{f-8}b). The ratio of $H_{c3}/H_{c2}\simeq 1.9$
at 8.5 K, this can be a result of either the enhanced $T_c$ by 0.13
K ($dT_c=0.2$) at the surface layer with the thickness of
$L/\lambda=0.71$, or, by assuming $b=-0.15/\lambda$. Decreasing the
temperature results in an increase of both: $L/\lambda$ and the
$H_{c3}/H_{c2}$ ratio. At $T=4.5$~K, $dT_c=0.028$
 and from Fig.~\ref{f-8}a, one finds that the value
of the $H_{c3}/H_{c2}$ ratio cannot exceed 1.75. Therefore, if the
GL theory is applicable, than, at $T=4.5$~K, the growth of the
$H_{c3}/H_{c2}$ ratio to 2.34 in Nb crystal is due to $b\neq 0$, and
the absolute value of $b$ increases with decreasing temperature.

The behavior of the ZrB$_{12}$ crystal is similar. The
$H_{c3}/H_{c2}$ ratio increases from 1.5 at 5.5~K to 1.75 at 4.5~K.
Since the ratio at 5.5~K is smaller than 1.69, it is possible that
$T_c$ of the surface layer is smaller than that of the bulk. Our
calculation shows that this value (1.5) can be obtained, for
example, if $dT_c=0.16$ and $L/\lambda=2$. For $T=4.5$~K the
dimensionless $dT_c$ is equals to 0.05 and $H_{c3}/H_{c2}$ ratio
does not exceed 1.65. We notice that in order to explain the
$H_{c3}/H_{c2}$ ratio, one has to take into account the nonzero
slope of the wave function at the surface.

\begin{figure}
     \begin{center}
    \leavevmode
       \includegraphics[width=0.9\linewidth]{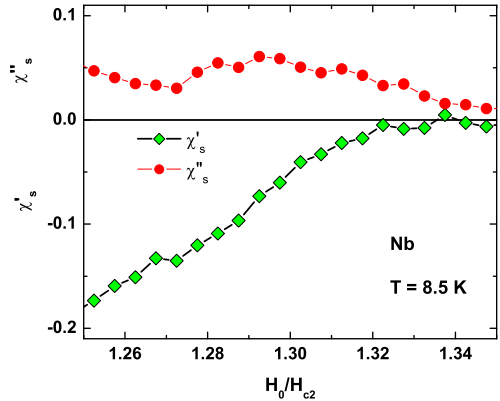}
    \caption{(Color online) Real and imaginary parts of surface susceptibilities  versus
    the dc magnetic field for Nb at 20 Hz and $T=8.5$~K near the point where the absolute
    value of both components are equal to each other ($H_0/H_{c2}\approx 1.3$).}

     \label{f-9}
     \end{center}
     \end{figure}

In general, with decreasing the dc field from its maximal value,
both the real and imaginary parts of the surface current appear
simultaneously, but at the beginning the imaginary part increases
faster than the real one. On the other hand, at $H_{c2}$ a complete
screening takes place and the absolute value of $\chi_{s}^{'}$
reaches its maximal value, $1/4\pi$, while $\chi_{s}^{''}=0$. So at
some dc magnetic field $\mid\chi_{s}^{'}\mid$ will be equal to
$\mid\chi_{s}^{''}\mid$. This point was identified in Ref.
~\cite{JUR} as the percolation transition from noncoherent SSS to
the coherent one in the Nb sample. For our Nb crystal, at $T=8.5$~K
and frequency 20~Hz it occurs at $H_0/H_{c3}\approx 0.68$, a value
which is slightly smaller than the value 0.81 reported
in~\cite{JUR}. However, our data presented in Fig.~\ref{f-9} does
not permit us to consider this point as a phase transition, due to a
smooth maximum $\mid\chi_{s}^{''}\mid$ at this field and the absence
of any peculiarity in $\chi_{s}^{'}$.

\begin{figure}
     \begin{center}
    \leavevmode
       \includegraphics[width=0.9\linewidth]{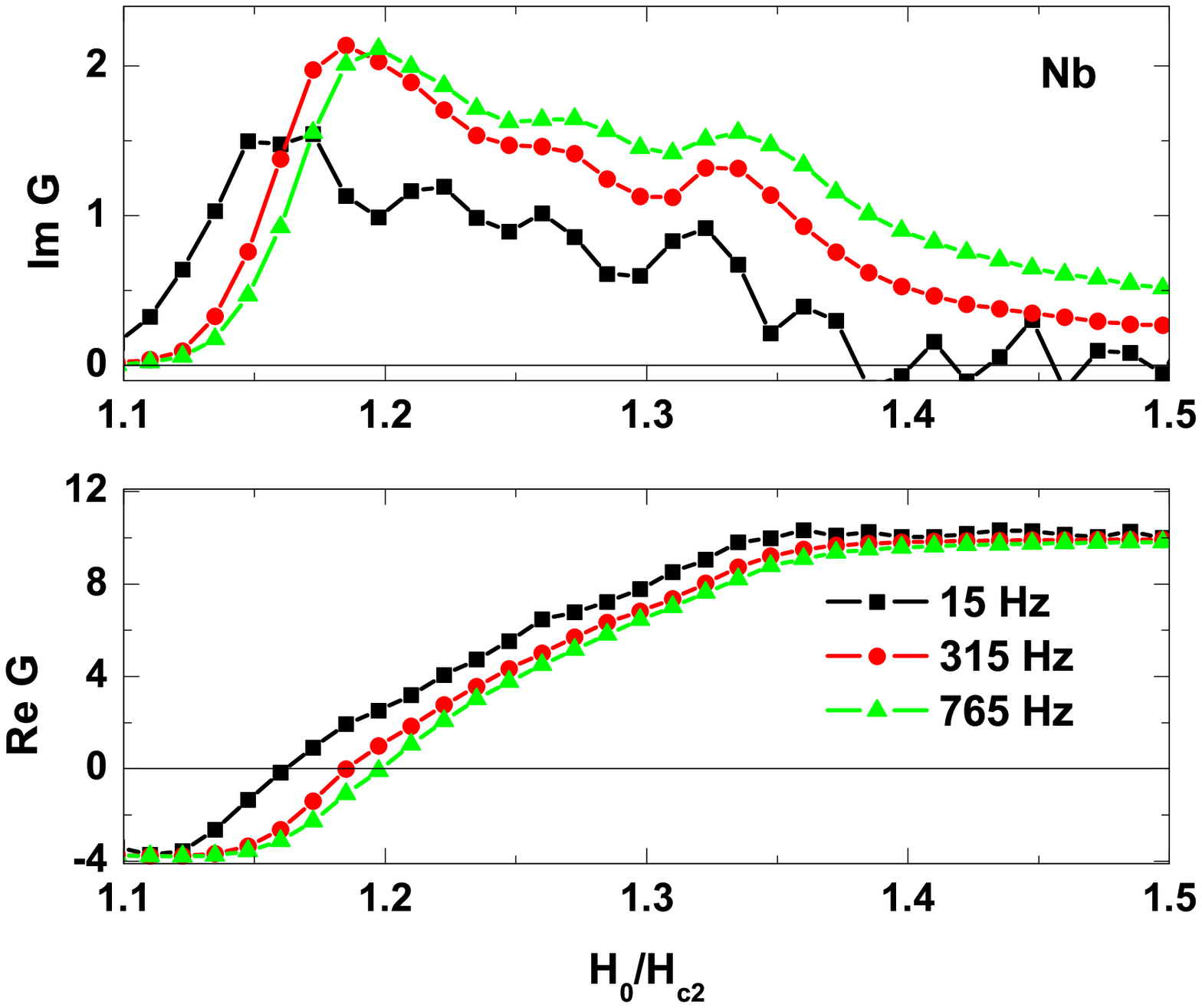}
        \includegraphics[width=0.9\linewidth]{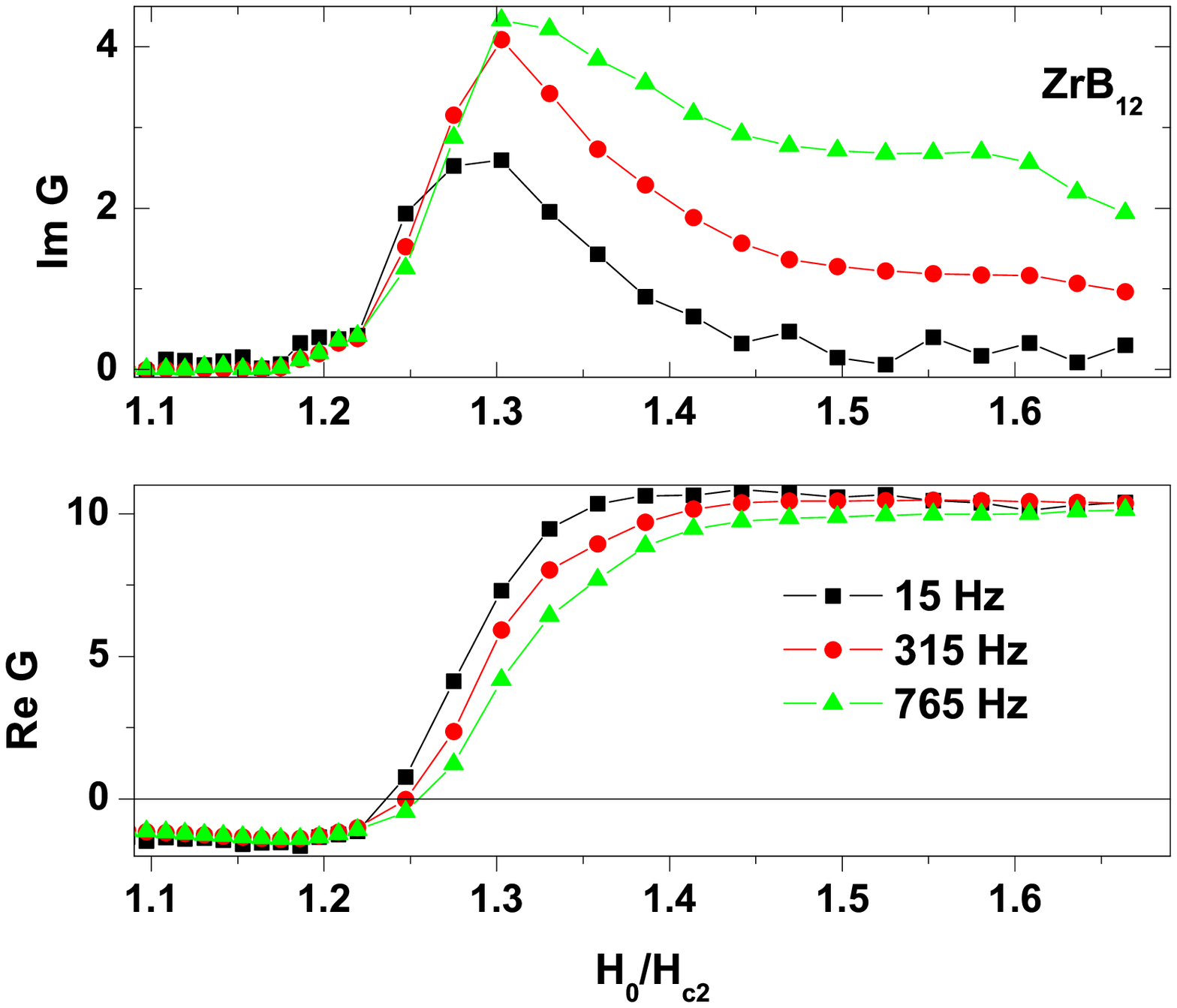}
    \caption{(Color online)The response function $G(\omega,H_0)$ of Nb and ZrB$_{12}$
    samples (upper and lower panels, respectively) versus reduced magnetic field, $H_0/H_{c2}$.  }

     \label{f-10}
     \end{center}
     \end{figure}

\begin{figure}
     \begin{center}
    \leavevmode
       \includegraphics[width=0.9\linewidth]{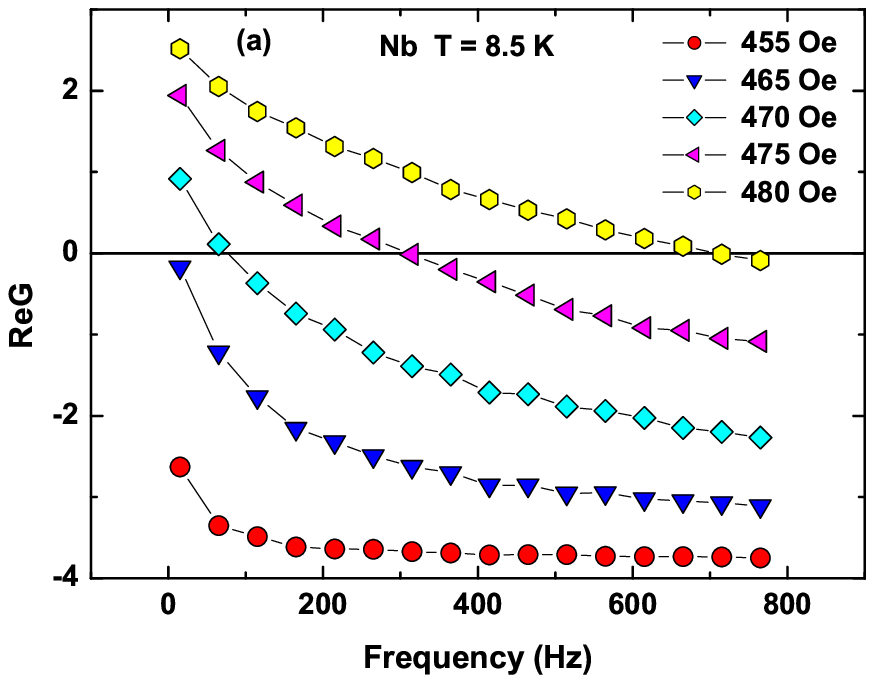}
       \includegraphics[width=0.9\linewidth]{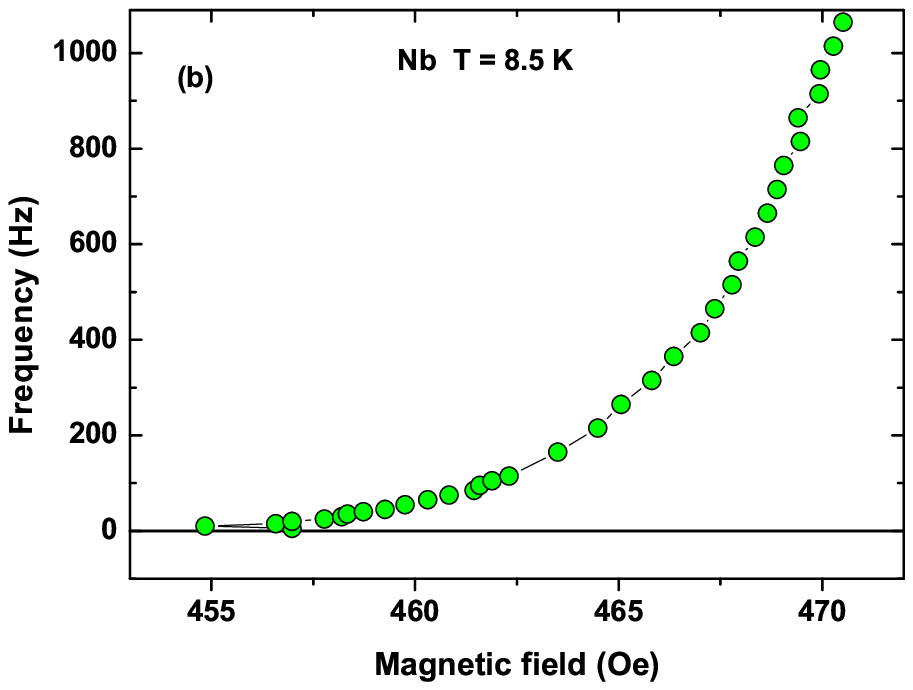}
    \caption{(Color online) (a) Frequency dependence of $G^{''}$ at different magnetic fields
    near the absorbtion maximum for Nb crystal at $T=8.5$ K.\\
    (b) Dispersion relation of the surface mode, $\omega_0(H_0)=\surd(\nu/\beta$ for Nb
    sample at $T = 8.5$ K }

     \label{f-11}
     \end{center}
     \end{figure}

The response of $K$ to an ac field is described by the function
$G(\omega,H_0)$ which can be found from Eq.~(\ref{Eq6}).
 Figs.~\ref{f-10}a,~\ref{f-10}b show $G(\omega,H_0)$ as
a function of a reduced magnetic field at several frequencies.
 Generally, $G(\omega,H_0)$ depends on the frequency.
Moreover, the real part of $G^{'}(\omega,H_0)$, changes its sign
with increasing the dc field. The frequency dependence of
$G^{'}(\omega,H_0)$ of the Nb crystal for several magnetic fields
near the loss peak, presented in Fig.~\ref{f-10} is shown in detail
in Fig.~\ref{f-11}a.
For some magnetic fields $G^{'}(\omega,H_0)$ changes its sign as the
frequency increases. The zero in the real part of the response
function can be considered as the manifestation of the existence of
unknown collective mode in the system. In Fig.~\ref{f-11}b we show
the frequency of this hypothetical surface mode as a function of the
dc field for Nb at 8.5K. The dynamics of $K$ cannot be described by
the first order linear differential equation:

\begin{equation}\label{Eq8}
  \partial K/\partial t=-\nu(K-\gamma h(t))
\end{equation}
with some relaxation constant $\nu$ and $\gamma =-
\frac{\frac{\partial J(H_0,K_0)}{\partial H_0}}{\frac{\partial
J(H_0,K_0)}{\partial K_0}}>0$. Equation~(\ref{Eq8}) gives the real
part $G^{'}(\omega,H_0)=\gamma\nu^2/(\nu^2+\omega^2)$, which does
not change its sign with frequency. One could expect that the second
order differential equation will give adequate description of the
observed response. But our experimental data show, that
$G^{'}(\omega,H_0)$ cannot be obtained from a differential equation
of comparatively low order.

The observed logarithmic frequency dependence of $\chi^{'}$,
Fig.~\ref{f-3}a for some dc fields, resembles the response of a
spin-glass system. But in spin-glass materials $\chi^{''}$ is a slow
function of the frequency as compared to $\chi^{'}$. Our data show
that the $\chi^{'}$ and $\chi^{''}$ values, both depend on the
frequency. Probably, similar to the spin-glass systems, we have here
a lot of clusters which are governed by the second order
differential equation, thus the observed response is the average
over all clusters. The second order differential equation
\begin{equation}\label{Eq8a}
    \partial K/\partial t=-\nu(K-\gamma h(t))-\beta
    \partial^2K/\partial t^2
\end{equation}
with $\beta >0$ has a response that changes its sign at
$\omega_0^2=\nu/\beta$. Assuming that $\nu/\beta$ for all clusters
increases with the dc field, we find, that the dc field value for
which $G^{'}(\omega,H_0)=0$ must increase with the frequency, as
observed in Fig.~\ref{f-10}. The $\sqrt{\nu/\beta}$ quantity in this
case will be the frequency of the surface collective mode.  The
dispersion relation of this mode, $\omega_0(H_0)$, for Nb is shown
in Fig.~\ref{f-11}b.

\begin{figure}
     \begin{center}
    \leavevmode
       \includegraphics[width=0.9\linewidth]{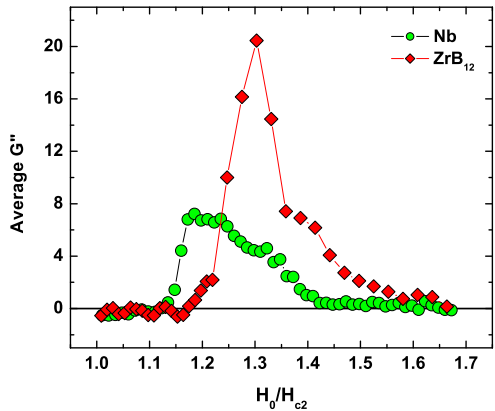}
    \caption{(Color online)  $\overline{G^{''}(\omega,H_0)}$
    over the frequency interval $0\leq\omega/2\pi\leq 20$ Hz as
    a function of reduced magnetic field, $H_0/H_{c2}$,
    for Nb at 8.5 K and ZrB$_{12}$ at 4.5 K.}

     \label{f-12}
     \end{center}
     \end{figure}

Further insight into the low-frequency response can be obtained from
the Kramers-Kronig relation~\cite{LL}
\begin{equation}\label{Eq10}
   \chi_{s}^{'}(\omega)-\chi_{\infty}=\frac{2}{\pi}\int^{\infty}_0\frac{\xi\chi_{s}^{''}
   (\xi)d\xi}{\xi^2-\omega^2}
\end{equation}
where $\chi_{\infty}=\chi_s(\infty)$. If we chose $\xi_0<<\omega<<\xi_m$ then:

\begin{equation}\label{Eq11}
\begin{array}{c}
\chi_{s}^{'}(\omega)-\frac{2}{\pi}\int^{\xi_m}_{\xi_0}\frac{\xi\chi_{s}^{''}(\xi)d\xi}{\xi^2-\omega^2}=\\
\bigskip

\chi_{\infty}+\frac{2}{\pi}(-\int^{\xi_0}_0\frac{\xi\chi^{''}(\xi)d\xi}{\omega^2}+\int^{\infty}_{\xi_m}
\frac{\xi\chi_{s}^{''}(\xi)d\xi}{\xi^2})
\end{array}
\end{equation}
The left side of this equation can be extracted from the available
experimental data. By presenting it as a linear function of
$1/\omega^2$ one can obtain the averaged imaginary part at low
frequency, defined as:
\begin{equation}\label{Eq12}
   \overline{\chi_{s}^{''}}(\xi_0)=2\int^{\xi_0}_0\xi\chi_{s}^{''}(\xi)d\xi/\xi_{0}^2.
\end{equation}
In Fig.~\ref{f-12} we have shown the averaged imaginary part of
$\overline{G^{''}}(\xi_0)$ ($G^{''}\propto\chi^{''}_s$), obtained
for Nb at $T=8.5$~K, using $\xi_0=2\pi 20$~s$^{-1}$,$\xi_m=2\pi
1065$~s$^{-1}$. $\overline{G^{''}}(\xi_0)$ in Fig. 12, is
considerably larger than any value shown in Fig.~\ref{f-10}. So the
Kramers-Kronig relations predict the existence of large loss peak at
low frequencies. This prediction needs a further experimental
evidence.

The observed ac response of SSS has a very complex character.
Partially, it can be the result of non-homogeneity of the samples.
The $K$ parameter, as it was introduced by Eq.~(\ref{Eq1}),
corresponds to the wave function of the whole sample. In real
samples, due to inhomogeneity, $K$ is not the integral constant of
the GL equations and the exact wave function is the superposition of
states with different $K$. Those states can relax with different
relaxation times as in spin-glass systems~\cite{SG}, and therefore
we observe such complicated responses. In addition, the dynamics of
the surface state is not described by a first order simple
relaxation equation. As a result, we obtain the logarithmic
frequency dependence of the real part of the susceptibility, as in
spin-glass systems, but the imaginary part shows a maximum at some
frequency.

\section{Conclusions}

In this paper we have presented investigation of the linear ac
susceptibility of Nb and ZrB$_{12}$ single crystals in the surface
superconducting state. Losses in this state have a linear origin,
and the critical state model for the surface current does not apply
here. Similar to spin-glass systems (where finite losses at
considerably low frequencies exist), the real part of susceptibility
exhibits a logarithmic frequency dependence. But the out-of-phase
component has a frequency dispersion. This dispersion in SSS differs
from that of the spin-glass systems. We assume that the sample
surface presents a lot of superconducting clusters, which are
governed by second order differential equation, and the observed
response is an average over these clusters. The Kramers-Kronig
analysis of experimental data reveals huge absorption peak at low
frequencies. The response of $K$ to the ac magnetic field defined by
Eq.~(\ref{Eq1}), $G(\omega,H_0)$, has been measured.

\section{Acknowledgments}

This work was supported by the INTAS program under the project No.
2001-0617 and by the Klatchky foundation for superconductivity. We
wish to thank Professors M. Gitterman, B. Rosenstein, and B.Ya.
Shapiro for many helpful discussions.

\end{document}